\documentclass[letterpaper, 10 pt, conference]{ieeeconf}  % Comment this line out
\IEEEoverridecommandlockouts                              % This command is only
\usepackage{balance}  % to better equalize the last page
\usepackage{graphics} % for EPS, load graphicx instead
\usepackage{times}    % comment if you want LaTeX's default font
\usepackage{url}      % llt: nicely formatted URLs

\makeatletter
\def\url@leostyle{%
	\@ifundefined{selectfont}{\def\UrlFont{\sf}}{\def\UrlFont{\small\bf\ttfamily}}}
\makeatother
\usepackage[pdftex]{hyperref}
\usepackage{cite}
\usepackage{color}
\usepackage{float}
\usepackage{lipsum}
\usepackage{graphicx}

\usepackage{amsmath}
\usepackage{algorithmic}
\usepackage{array}

\begin{document}
\title{\LARGE \bf
	Cognitive Systems Approach to Smart Cities
}

%\author{\IEEEauthorblockN{Aladdin Ayesh}
%	\IEEEauthorblockA{Faculty of Technology, \\ De Montfort University\\
%		Leicester LE1 9BH, UK\\
%		Email: aayesh@dmu.ac.uk}}

\author{Aladdin Ayesh \\ {\small Faculty of Computing, Engineering and Media,}\\ {\small De Montfort University, UK} \\ {\tt\small aayesh@dmu.ac.uk}} %$^{*}$ % <-this % stops a space
%\thanks{*This work was not supported by any organization}% <-this % stops a space
%\thanks{$^{*}$ A. Ayesh is with Faculty of Technology, De Montfort University, UK
%{\tt\small aayesh@dmu.ac.uk}}%
%\thanks{$^{2}$P. Misra is with the Department of Electrical Engineering, Wright State University,
%       Dayton, OH 45435, USA
%       {\tt\small p.misra at ieee.org}}%

%}

\maketitle
\thispagestyle{empty}
\pagestyle{empty}

%%%%%%%%%%%%%%%%%%%%%%%%%%%%%%%%%%%%%%%%%%%%%%%%%%%%%%%%%%%%%%%%%%%%%%%%%%%%%%%%
\begin{abstract}

In our connected world, services are expected to be delivered at speed through multiple means with seamless communication. To put it in day to day conversational terms, “there is an app for it” attitude prevails. Several technologies are needed to meet this growing demand and indeed these technologies are being developed. The first noteworthy is Internet of Things (IoT), which is in itself coupled technologies to deliver seamless communication with “anywhere, anytime” as an underlying objective. The “anywhere, anytime” service delivery paradigm requires a new type of smart systems in developing these services with better capabilities to interact with the human user, such as personalisation, affect state recognition, etc. Here enters cognitive systems, where AI meets cognitive sciences (e.g. cognitive psychology, linguistics, social cognition, etc.). 

In this paper we will examine the requirements imposed by smart cities development, e.g. intelligent logistics, sensor networks and domestic appliances connectivity, data streams and media delivery, to mention but few. Then we will explore how cognitive systems can meet the challenges these requirements present to the development of new systems. Throughout our discussion here, examples from our recent and current projects will be given supplemented by examples from the literature.

\end{abstract}


%%%%%%%%%%%%%%%%%%%%%%%%%%%%%%%%%%%%%%%%%%%%%%%%%%%%%%%%%%%%%%%%%%%%%%%%%%%%%%%%
\section{INTRODUCTION}

One may argue that cities denote space and services. Space is occupied by inhabitants who use services which are either utilities, e.g. water, or transport. However, technological advances and the introduction of smart devices that enables inhabitants to expand on services and manipulate their environment, i.e. space, increased expectations of what is possible within a city. At the same time, global warming, limitation of resources, and economical considerations made urban planners rethink the concept of a city leading to the introduction of smart (sustainable) cities. In an ideal smart city, the whole city is connected. The delivery of services requires minimum human intervention and minimum wastage of resources. The inhabitants are always informed not just about services but also events. The events, whether happy events such as festivals or catastrophic ones such as hurricanes, are well managed because everyone is connected everywhere at anytime, and because services are well planned and optimally distributed though out the city. This idealistic view while is possible, requires the development of a new generation of smart systems that can seamlessly integrates in space and services that interacts with users, i.e. inhabitants, intuitively. At the same time, these systems should have the flexibility to adapt to users and expanding cities needs. In this paper we explore cognitive systems as a potential approach to setting up a platform for smart cities that can deliver services and communication with intuitive interaction that has the ability to adapt and expand. We start the paper by looking at smart cities, then cognitive systems leading to a sketch of proposal for smart city cognitive platform. We conclude with reflective evaluation. % leading to future work plans. 

\section{Smart Cities}

\subsection{Preliminaries}

There have been several attempt to describe, define or explain the concept of smart cities \cite{Falconer2012,Sta2017}. Some of these attempts focus on technology whilst others evolves from public policy or sustainability in urban planning. Olivereira and Painho \cite{Oliveira2015} advocates that a smart city has four primary dimensions of operation referencing the work of Roche \cite{Roche2014}. These four dimensions are: intelligent city, digital city, open city, and live city. Intelligent city represents the social infrastructure whilst digital city implies the information infrastructure. It is clear both are strongly connected and implies, in general, \textit{infrastructure}, perhaps with a focus on inhabitants of the smart city who establish or generate, utilise, and maintain social and information components of the city infrastructure with the aide of technological advances \cite{Roche2014}. Open city is a reference to open government. Finally live city denotes the urbanization of space that is continually adaptive to the living needs of its inhabitants. This last dimension is often the one at the forefront of smart cities thinking where issues such as sustainability and technological requirements are identified and extensively examined.  

This dimensional view in conjunction with the graphical representations provided by Pramanik \textit{et. al.} \cite{PRAMANIK2017370} and by Jucevicius \textit{et. al.} \cite{Jucevicius2014} led us to identify basic components of what may be meant by a smart city. We use these components in an attempt to define what constitutes a smart city. 

\subsection{To Define a Smart City}
\label{sec:smartcity}
\begin{figure}[H]
	\includegraphics[width=\linewidth]{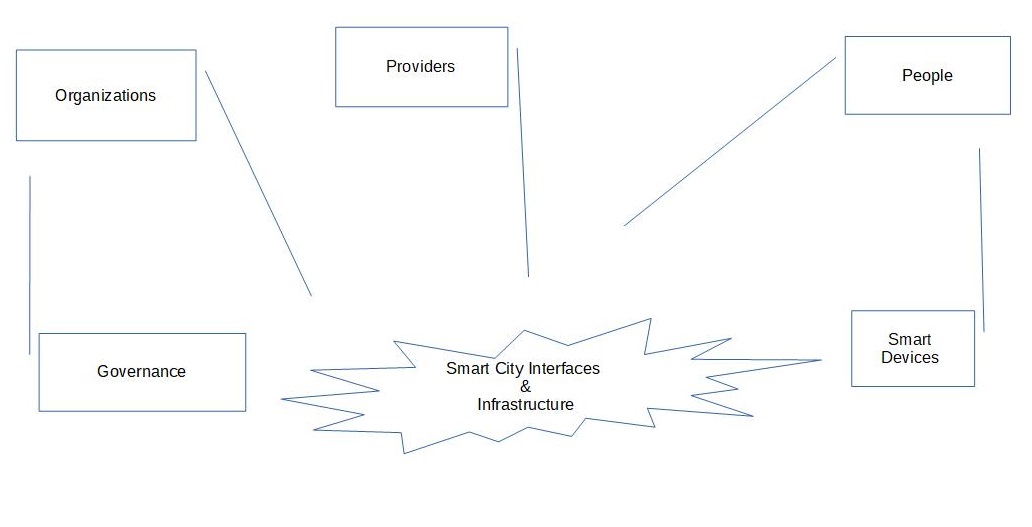}
	\caption{A City's Stakeholders}
	\label{stakeholders}
\end{figure}

It is worth establishing that our approach here to the concept of a (smart) city is a \textit{systems'} approach, although be it cognitive. Thus, we will be looking at smart city as a system that requires an architecture and modularisation, possibly a multi-dimensional one. The first two dimensions proposed by \cite{Oliveira2015} relates to infrastructure and inhabitants, thus it would be useful to identify the stakeholders in a city leading to some understanding of where and how the \textit{smart} label makes its way to be attached to a city. In doing so, we follow, as aforementioned, a systems' approach. Figure \ref{stakeholders} provides a visual representation of the major categories of stakeholders.

We can see that stakeholders do not come unaccompanied. They have enabling objects or smart devices, e.g. mobile phones, wearable trackers, etc., and they have functions such as the case in governing organizations. Providers category is large and extensive. It includes providers of technology, services, infrastructure, management, etc. In other words, there are individuals, companies, and governing institutes, all of which relies on the city infrastructure and interfaces. With technological advances, city inhabitants are expecting their smart devices to enable them to interact with smart infrastructure with seamless interfaces that are equipped to cope with human intelligence and to communicate in more natural and intuitive (smart) forms.

\begin{figure}[H]
	\includegraphics[width=\linewidth]{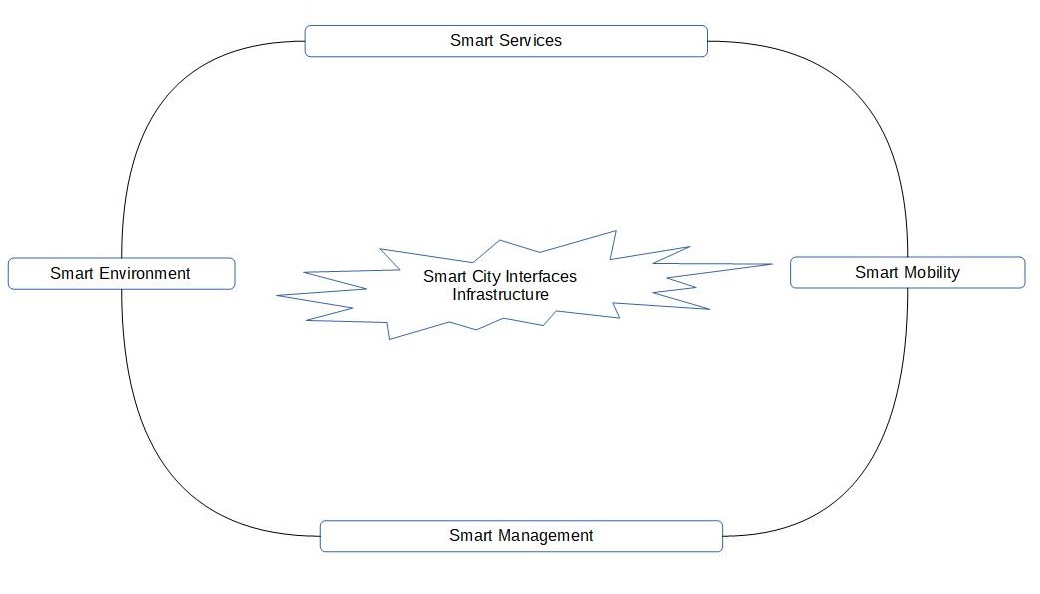}
	\caption{Major Components of a Smart City}
	\label{components}
\end{figure}

\subsection{Challenges}

The biggest challenge in developing and maintaining a smart city is the integration of all the components and their users \cite{Falconer2012,Combaneyre2015}. There have been several attempts to circumscribe these components into a discrete set with some degree of success \cite{PRAMANIK2017370, Jucevicius2014}. It is inevitable that any particular piece of research will focus on one facet of any prescribed set of components. We provide our own attempt here building on what have been reported in literature with two guiding thoughts: low granularity in modularisation and high relativity to cognitive computing. The low granularity allows us to categorise smaller components and to identify the major similarities between them that is necessary for a generalised platform. At the same time, it allows us to see the bigger picture through abstraction, which enables us to identify the architectural elements of such platform.         

\subsubsection{Smart Environment}
First of these umbrella components relates to space part of a city. Space here denotes both public and private, indoor and outdoor spaces. These spaces provide the environment occupied by inhabitants and services including the technology that enables both habitation of environment and delivery of services. 
 
\begin{figure}[H]
	\includegraphics[width=\linewidth]{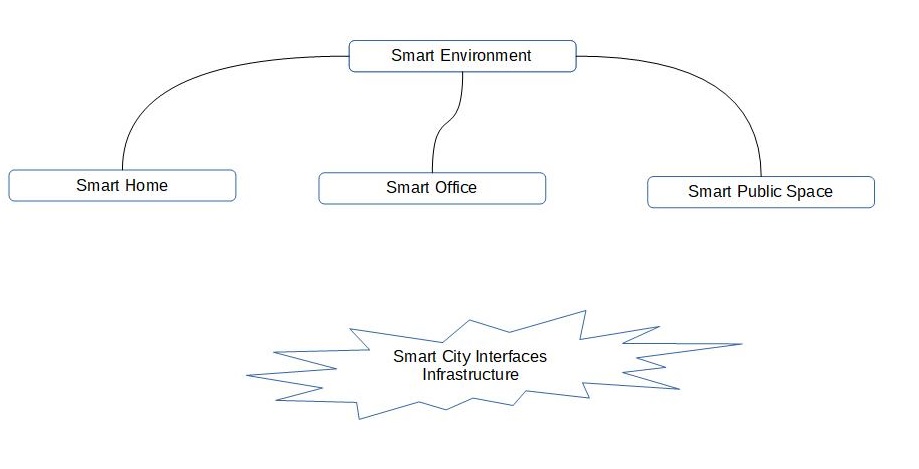}
	\caption{Smart Environment - Spaces}
	\label{spaces}
\end{figure}

Immediately we can criticise figure \ref{spaces} for ignoring virtual spaces. If we are to accommodate open city dimension with many of e-government services delivered in virtual spaces, this type of space must be included in our smart environment component of smart city. In addition, virtual spaces are not exclusively limited to governance dimension. The advances in virtual environment technologies, e.g. mobile phones virtual reality goggles, in conjunction with flexible work initiatives, to give an example, start to blur the boundaries between physical and virtual spaces. Thus while we do not include explicitly virtual spaces in figure \ref{spaces} it is to be understood virtual spaces are at the heart of smart environment and augment physical spaces of home, work and public.   

\subsubsection{Smart Services}
\begin{figure}[H]
	\includegraphics[width=\linewidth]{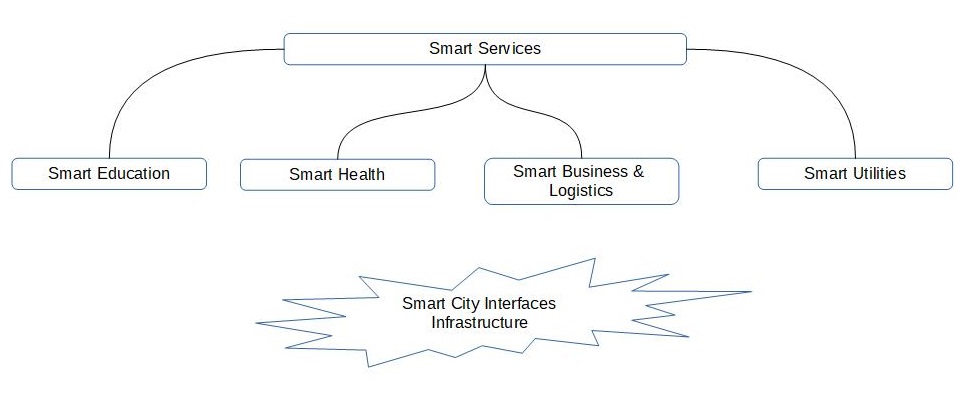}
	\caption{Smart Environment - Services}
	\label{services}
\end{figure}

A smart environment implies smart services that will be embedded in the designated space of such environment. For example, a smart building will incorporate into its fabric smart devices, such as smart meters, to enable relevant smart services such as smart utilities. In this particular example data may be collected on the setup of the house, such as optimum or desirable temperature, appliances use, time and duration of use or peak times, etc. Another example is a smart road which would also incorporate devices such as speed cameras, communication beacons, information screens, etc. Figure \ref{services} summarises key smart services that are to be expected in a smart city. 

\subsubsection{Mobility and Management}
From a system view, we can look at Mobility as a control problem. Similarly we can view Management as a constrain satisfaction problem with governance policies forming the constraints. Both mobility and management are effectively decision making problems constrained by resources planning. This reductionist approach by no means trivialise the long list of challenges mobility and management imposes on technology development, e.g. hardware, and on other disciplines, e.g. public policy research \cite{KOURTIT201713,LYONS2016,Komninos2011}.    

\subsection{Big Data and Sustainability}

In developing smart cities platform(s) two major issues cause the majority of technological challenges. These are Data \cite{Sta2017,KOURTIT201713,Ayesh2014e,Batty2013} and Sustainability \cite{BIBRI2017183,LYONS2016,Obaidy2015,Bertelle2010}. In a fully connected city, data is generated from every smart device being a mobile phone, a smart electricity meter, or any other device. This data needs to be processed, stored fully or partially, analysed and feedback into decision systems for example to recommend a service or to adapt the environment. In the context of smart cities, the data challenges is not just in the volume but in what is known as big data big V's \cite{PRAMANIK2017370}.    

Maintaining the underlying technology of smart city is a great challenge for sustainability research \cite{BHATI2017230,Anthopoulos2016,BIBRI2017183,Bertelle2010}. 
%\subsection{Infrastructure Sustainability}

%Dynamic Infrastructure and Sustainability are two interlocked topics of most importance for smart cities research \cite{BIBRI2017183,LYONS2016}. In fact, smart cities are so often linked with sustainability that most research on smart cities would address sustainability as inseparable part of what constitutes a smart city. 

\section{Cognitive Systems}

\subsection{Components}
It is difficult to specify fixed components for cognitive systems. They often rely on cognitive architectures \cite{Langley2006}. Each architecture often emphasises one or few cognitive capabilities depending on the problem for which the architecture is designed to solve. Langley \textit{et. al.} \cite{Langley2009} provide interesting review of cognitive architectures in general. They give a useful summary of the expected capabilities in cognitive architectures, namely: Recognition and categorization, Decision making and choice, Perception and situation assessment, Prediction and monitoring, Problem solving and planning, Reasoning and belief maintenance, Execution and action, Interaction and communication, and finally, Remembering, reflection, and learning. Equally, their specified properties of cognitive architectures are of equal interest: Representation of knowledge, Organization of knowledge, Utilization of knowledge, and Acquisition and refinement of knowledge. 

\subsection{Limitations}
There are three major limitations we can cite for cognitive systems from analysing the literature. First, cognitive systems as they imitate human cognition will suffer from the same limitations of the human cognitive system, e.g. dealing with incomplete or ambiguous information.

Second, there is the limitations of the enabling technology, e.g. vision systems, processing power, etc. It may be easy to specify a cognitive system in terms of hardware and software, however, the practical capabilities during real life deployment scenario will be greatly challenged by the limitations of the hardware, e.g. consider the speed and scope of communication in a human nerve system. This may be referred to as an integration problem, which is one of the important topics in cognitive systems and the development of cognitive architectures \cite{Chong2007}.

Third limitation is in evaluation mechanisms. We have to bear in mind that there is an element of subjectivity in the user interaction with a cognitive system, consider as an example different users reactions to recommendations by a recommender system. This is a general AI problem. A quick review of the literature related to the Turing Test is sufficient to demonstrate the difficulty and complexity of testing a cognitive system that integrates several AI capabilities. That did not deter researchers from pursuing the task of identifying evaluation criteria. A good example of system approach can be seen in \cite{Langley2009}.

%\subsection{In the Context of Smart Cities}

\section{Towards a Cognitive Platform}

\subsection{Requirements}
 
Table \ref{requirements} summarises some key requirements if we attempt to develop a general cognitive platform for a smart city. The core capabilities, namely cognitive, are the obvious ones and yet the more complex to attain in full. In most practical cases a concrete substantiation of the platform will implement a subset of these capabilities.  The other capabilities, namely secure and sustainable, are essential, we may even argue unmissable in any implementation of smart city cognitive system. It is worth highlighting that some of the cognitive capabilities of the system are likely to be used to provide the capabilities of being secure and sustainable, e.g. by using learning and prediction capabilities \cite{Muhammad2017,Abdelhamid2014,Obaidy2015,Alruily2014,Daoud2014}.

\begin{table} [htbp]
	%\twocolumn[
	\caption{Summary of Requirements}
	\begin{tabular}{|p{2.5cm}|p{1.5cm}|p{1.5cm}|p{1.5cm}|}
		\hline
		\textbf{Capabilities} & \textbf{Technologies} & \textbf{Analytics} & \textbf{Examples} \\ \hline
		Mobile, Multi-Sensors, Heterogeneous content & Context aware, intelligent agents, mobile networking protocols & cognitive and sensor networks analytics & Driverless cars \\ \hline
		% &  &  &  \\ \hline
		&  &  &  \\ \hline
		Cognitive Capabilities:	&  &  &  \\ \hline
		Knowledge management & Big data storage and management, e.g. Apache Hadoop \& Solr & Cognitive analytics  & Location-based services \\ \hline
		Decision making & Big data processing, e.g. ReduceMap & Social network analysis  & Recommender systems \\ \hline
		Interaction, Communication & Machine learning and intelligent agents & Sentiment analysis & Personalisation \\ \hline
		&  &  &  \\ \hline
		Secure, Privacy & Machine learning  & Visualisation & Privacy tolerance recognition  \\ \hline
		&  &  &  \\ \hline
		Sustainable & Intelligent agents & Visualisation & Self-diagnosis and maintenance,  \\ 
		&  &  & energy harvesting \\ \hline
	\end{tabular}
	\label{requirements}
	%	]
\end{table}

Cognitive capabilities require special attention because of their variety in functionality, technology used and applications. Thus we will look at these capabilities within the context of some on going relevant research.

\subsection{Recommender Systems}
Recommender systems are effectively data-based decision systems. They fuse data gathered from variety of sources, e.g. home sensors, social media streams, etc. They turn the fused data into subject knowledge, e.g. user or location profile. They feature heavily in online shopping platforms, however, their application and potential goes beyond recommending to customers the next product to buy. 

Location-based services (LBS) is one example in wide use by smart devices from car satellite navigators to smart sport watches. LBS can still be improved as a technology \cite{Daoud2014} but the real advances is likely to come from cognition where users' cognition factors, e.g. perceptions, social networking, sentiments, intention, etc., are taken into consideration \cite{Alruily2014,DAniello2016}. By injecting cognitive capabilities in recommender systems they can become the controlling hub in the embedded systems of smart environments integrating smart environment components, e.g. sensors, whilst managing smart services provided to inhabitants based on variety of criteria such as location and user profiles. This can also facilitate other aspects of running the smart city such as relaying information in case of emergencies with personalised recommended actions, to give an example.

\subsection{Communication and Governance}
\label{com}
Smart environments are occupied by inhabitants with whom a cognitive recommender system that oversees the smart environment needs to interact and communicate. This communication in a living city does not always have to be on an individual basis nor should be assumed it will always be a direct communication. This inevitably leads us to issues of governance and public policy. Crowd management is a popular research topic and a good example \cite{8000564,KOUSIOURIS2017,KUMAR2017}. It furnishes us with use case scenarios in providing services \cite{Abdelhak2012,7390614}, managing events \cite{KOUSIOURIS2017} and their large data streams, and in providing security \cite{KUMAR2017}. Now if our recommender system allows for public policies to be represented within its internal representation in a fashion similar to what Niklaev and Ayesh did \cite{Nikolaev2011a} technologies such as HTN planners can be utilised producing personalised crowd management plans as recommendation. Why should it be personalised? This is to allow the system to take into consideration and balance between the localised and individual circumstances versus the larger triggering event. 

%As an example, let us look at an example of a hurricane. The general advise is to evacuate. Now there is a mother with a baby in the path of the hurricane. They should evacuate, however, there i 

\subsection{Services Distribution and Logistics}
Service management in part relies on planning and optimal distribution of services around the city 
\cite{Bertelle2010, Assem2016,BIBRI2017183,KOURTIT201713}. Accessibility of these services often rely on logistics and mobility \cite{LYONS2016}. Here is where the question of sustainability appears strongly. On one hand resources, such as energy, need to be managed. On the other hand more services require more resources. Obaidy and Ayesh \cite{Obaidy2015} provides an example where a balance between energy consumption and coverage area maximization is required, in this case the example is from mobile sensor networks with potential applications in security surveillance and crowd management amongst others relating this to what we have covered in section \ref{com}.

\section{Reflective Evaluation}

In reviewing literature and analysing our current understanding of smart city, a cognitive platform is both feasible and desirable. We can already see components of this platform being developed in a standalone systems or as subsystems of related contexts. However, a standard in developing these components or subsystems is necessary to enable their integration into the larger, more dynamic and heterogeneous system that is a smart city. 

We identified two major umbrellas out of the four parts forming a smart city where cognitive systems can make substantial contribution, namely smart environments and smart services. Both are characterised by being dynamic and evolving whilst requiring direct interaction with the inhabitants where cognitive abilities would be an advantage.

\section{CONCLUSIONS}
In this paper we attempted to examine cognitive systems approach to the technological needs of a smart city. In the process we identified some of the requirements of developing a cognitive platform for smart cities and initial specifications of implementing some of the subsystems such a platform requires.

%\addtolength{\textheight}{-8cm}   % This command serves to balance the column lengths

%%%%%%%%%%%%%%%%%%%%%%%%%%%%%%%%%%%%%%%%%%%%%%%%%%%%%%%%%%%%%%%%%%%%%%%%%%%%%%%%

%%%%%%%%%%%%%%%%%%%%%%%%%%%%%%%%%%%%%%%%%%%%%%%%%%%%%%%%%%%%%%%%%%%%%%%%%%%%%%%%

%%%%%%%%%%%%%%%%%%%%%%%%%%%%%%%%%%%%%%%%%%%%%%%%%%%%%%%%%%%%%%%%%%%%%%%%%%%%%%%%

%\section*{APPENDIX}


%\section*{ACKNOWLEDGMENT}

%%%%%%%%%%%%%%%%%%%%%%%%%%%%%%%%%%%%%%%%%%%%%%%%%%%%%%%%%%%%%%%%%%%%%%%%%%%%%%%%

\bibliographystyle{IEEEtran}
\bibliography{References/smartCities}

\end{document}